\documentclass[aps,pra,twocolumn,superscriptaddress,longbibliography]{revtex4-2}
\usepackage{graphicx}
\usepackage{amsmath,amssymb}
\usepackage{hyperref}
\graphicspath{{./}}

\begin{document}

\title{Information capacity of quantum statistics: Fock-state tests of
a discrete binary-sequence model on cloud photonic quantum processors}

\author{Chiran Wijesundara}
\author{Octavia T. Volpe}
\author{Dejan Stojkovic}
\author{Herbert Fotso}
\author{Tim Thomay}
\affiliation{Department of Physics, SUNY at Buffalo, Buffalo, NY 14260-1500, USA}

\begin{abstract}
The central premise of this work is that quantum mechanics may be the statistical limit of a more fundamental discrete theory.
Any such theory equips a physical system with a finite information capacity, and its departure from quantum statistics is controlled by how much of that capacity the system uses.
We show that commercial cloud photonic quantum processors have reached the precision required to bound this capacity from below, using the binary-sequence model of Powers \emph{et al.} as the concrete, falsifiable theory that makes the capacity operational: outcome probabilities arise from counting discrete sequences of length $n$, quantum mechanics is recovered as $n \to \infty$, and $n$ measures the information capacity of the register behind a prepared state.
Photon Fock states $|1\rangle$, $|1,1\rangle$, heralded $|2\rangle$, and cascaded beam-splitter pairs are measured on programmable interferometers and compared against the quantum-mechanical partition law and the discrete model at finite $n$.
All dominant systematics are determined in situ and treated as joint or marginalized nuisance parameters: transmittance calibration, a source indistinguishability of $0.85$, output-port efficiency asymmetry, source stability, and a frozen per-circuit compilation offset of $0.026$ in transmittance.
The model's composition-consistent parametrization, singled out by requiring that rotations compose, is its unique formulation with a quantum limit: it recovers quantum mechanics as $n \to \infty$ with deviations $1.24/n$.
For this model a random-effects likelihood analysis calibrated by parametric bootstrap under both hypotheses excludes all $n \le 100$: the information capacity of the register carrying the two-photon state, if finite, exceeds $10^{2}$.
Cascaded beam splitters test the model's composition law directly: the data are split-invariant, excluding naive count composition at $8\sigma$ in a calibration-free differential test and confirming the interference-sign composition rule.
Model-independently, curve-averaged coherent deviations from the quantum partition law in the $P(1,1)$ channel larger than $2.3\times10^{-2}$ are excluded at 95\% confidence.
The originally published parametrization, linear in the sequence counts, lacks a quantum limit and is excluded outright: its unconditioned variant for all sequence lengths and its conditioned variant for $n = 32$ to $400$.
Because the compilation offset is frozen per circuit it is calibratable, opening the $10^{-3}$ statistical floor ($n \sim 10^{3}$) to current cloud hardware.
These results establish cloud photonic processors as quantitative instruments for quantum foundations, and the information capacity of a quantum system as an experimentally boundable quantity.
\end{abstract}

\maketitle

\section{Introduction}
\label{sec:intro}

Quantum mechanics is the most precisely tested framework in physics, yet the question of whether its probabilistic structure is fundamental or emergent remains open, and a growing experimental program treats it as such: precision bounds on higher-order (Sorkin) interference
\cite{Sinha2010,Kauten2017,Pleinert2021} and on deviations within the
landscape of generalized probabilistic theories \cite{Mazurek2021} constrain the very structure of quantum probability, and cloud quantum processors have begun to serve as foundations testbeds \cite{Ku2020,Santini2022}.
A recent theory by Powers \emph{et al.}
\cite{Powers2022,Powers2025} constructs quantum states
from ensembles of finite binary sequences, in the spirit of Wheeler's ``it from bit'' program \cite{Wheeler1990}: probabilities arise from counting discrete ontic configurations of sequence length $n$, and standard quantum mechanics (QM) is recovered in a large-$n$ limit.
Among other things, the model successfully recovers the quantum mechanical rules for angular momentum composition and Clebsch-Gordan coefficients \cite{Powers2022}, the Wigner $d$-matrix formula \cite{Powers2025}, and the quantum harmonic oscillator \cite{OscillatorTBD2026}.
The standard quantum non-determinism is a consequence of obscuring the exact composition of a sequence while retaining only the quantum numbers that count how many times a certain basis element appears in a sequence.
Quantum probabilities are reduced to counting sequences, while quantum interference is driven by non-local quantum numbers.
Thus, this informational theory framework appears capable of answering many questions that could not be answered in the context of quantum mechanics itself.

At finite $n$ the model makes falsifiable predictions while respecting the Bell \cite{Bell1964}, Kochen--Specker \cite{KochenSpecker1967}, and PBR \cite{PBR2012} no-go theorems: percent-level deviations from the quantum probabilities of photon-number partition experiments, and outcomes that are exactly forbidden where QM assigns small non-zero probability.
The theory joins a family of discrete and emergent reconstructions of quantum theory, from spin networks
\cite{Penrose1971} to cellular-automaton \cite{Hooft2016} and
finite-field \cite{Chang2013} formulations, but is distinguished by an event-network construction \cite{Powers2023event} that, in its beam-splitter form \cite{Powers2025}, yields concrete, platform-ready predictions.
Testing such a theory does not require overthrowing quantum mechanics; a null result converts directly into a lower bound on the discreteness scale $n$.
This scale has a direct physical reading as an \emph{information capacity} \cite{OscillatorTBD2026,StojkovicNote2026}: any finite system carries finite information, and the experiments below measure how much of it a photon pair must at least have.
That reading places the test in the lineage of informational reconstructions of quantum theory \cite{Zeilinger1999,Chiribella2011} and of information-theoretic principles that bound quantum correlations \cite{Pawlowski2009}.
The model plays the role here that a test theory plays in searches for Lorentz violation: it turns the open question of whether quantum mechanics is exact into a measurable parameter.
The parameter it exposes is the capacity of the underlying register, not the Holevo capacity \cite{Holevo1973} of the two-photon state, which is fixed at $\log_2 3$ bits by Hilbert-space dimension alone and which no experiment can change.

The natural testbed, proposed alongside the model itself in Ref.~\cite{Powers2025}, is a variable beam splitter fed with photon-number (Fock) states: the model's predictions for the output partition probabilities are closed-form, the quantum prediction is the elementary binomial law \cite{Campos1989,Loudon2000,Gerry2005}, and the required hardware, namely programmable interferometers with heralded single-photon inputs and photon-counting readout, is now commercially available as cloud-accessible photonic quantum processors.
This paper reports the first systematic campaign of this kind, with three contributions.
First, a complete, open experimental protocol: circuit constructions compatible with hardware constraints (heralded two-photon preparation, photon-number-resolving readout verification), an in-situ nuisance chain that measures every dominant systematic from the same data that carry the physics, and an information-theoretic experiment-design step that closes the model's internal escape routes.
Second, quantitative constraints: for the composition-consistent parametrization, under which the model uniquely recovers QM, a bootstrap-calibrated random-effects analysis bounds the information capacity to $n > 100$; a cascaded-beam-splitter campaign excludes the naive composition law ($8\sigma$ jointly, calibration-free differential test) in favor of the interference-sign rule; we bound the curve-averaged coherent deviation from the quantum partition law in the $P(1,1)$ channel below $2.3\times10^{-2}$ at 95\% confidence; and the published linear-parametrization variants are excluded outright on two processor generations (Quandela Ascella and Belenos: unconditioned for all $n$, conditioned for $n = 32$ to $400$).
Third, a sensitivity benchmark: today's photonic cloud platforms are noisy intermediate-scale quantum (NISQ) devices, and we identify and quantify their systematics ladder (compilation-to-compilation transmittance jitter, source stability, port-efficiency asymmetry, multiphoton contamination) and state which future bounds each rung permits.
The entire hardware campaign consumed only a modest amount of cloud machine time, so precision foundations tests of this kind are accessible to essentially any research group.

\section{Model predictions for beam-splitter partitions}
\label{sec:model}

The model of Ref.~\cite{Powers2025} replaces continuous transition amplitudes by a finite collection of sequences representing the underlying ontic configurations of the system.
An observer never has access to the full microscopic sequence, only to how often each symbol appears in it: these symbol counts play the role of quantum numbers, and the loss of the detailed ordering information is what restores quantum indeterminism.
Concretely, a measurement event pair is represented by base-16 sequences of length $n$, and outcome probabilities are obtained from an alternating sum over contextual-set cardinalities [Eq.~(20) of Ref.~\cite{Powers2025}, evaluated from the counts of its Table~II].
For photon-number states on a beam splitter, the input Fock state $|N_1, N_2\rangle$ maps to a spin system with $j = (N_1{+}N_2)/2$, $m_{a1} = (N_1{-}N_2)/2$, and the rotation map between the event pair carries $\widetilde{B} \in \{0,\dots,n\}$ flip symbols (and $\widetilde{A} = n - \widetilde{B}$ identity symbols): at finite $n$ the transmittance lives on a rational grid.
We implemented the count machinery independently from the published tables and verified it against the authors' reference code \cite{PowersCode} to machine precision ($7\times10^{-16}$ across integer and half-integer $j$, conditioned and unconditioned internal quantum numbers), and against the published figures.

\begin{figure}[t]
\includegraphics[width=\columnwidth]{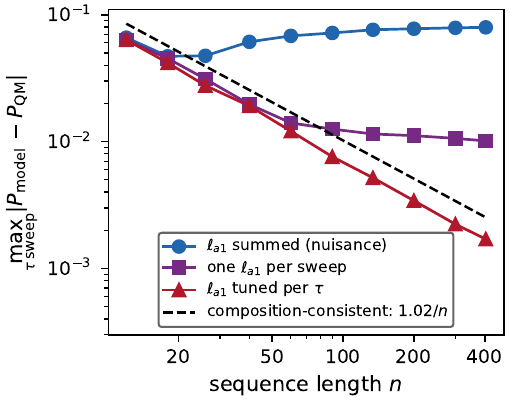}
\caption{Maximum partition-probability deviation from QM against sequence
length $n$.
Published parametrization ($|2,0\rangle$ input, mid-$\tau$ sweep): $\ell_{a1}$ summed (plateau), one $\ell_{a1}$ per sweep ($1/n$ to a $\sim$1\% floor), and $\ell_{a1}$ tuned per setting (clean $1/n$).
The composition-consistent parametrization (black dashed) makes a single prediction, $1.02/n$ for $|2,0\rangle$ over the full grid.}
\label{fig:scaling}
\end{figure}

The dictionary between the map count and the optical transmittance $\tau = \cos^2(\theta_{ab}/2)$ involves a calibration choice.
Ref.~\cite{Powers2025} fixed it at the level of a single rotation, $\theta_{ab} = \widetilde{B}\pi/n$, the simplest assignment matching the Born rule setting by setting.
This original assignment, linear in the count $\widetilde{B}$, is what we call the published parametrization throughout.
Extending the model to
\emph{composed} rotations singles out a refinement of this dictionary,
$\tan(\theta_{ab}/2) = \widetilde{B}/\widetilde{A}$
\cite{StojkovicNote2026}: the map is characterized by the complex
number $z = \widetilde{A} + i\widetilde{B}$ with $\arg z = \theta_{ab}/2$, composition of rotations becomes complex multiplication, and angles add exactly.
The refinement also sharpens the large-$n$ limit.
Under the composition-consistent calibration the model, summed over its internal quantum number $\ell_{a1}$, converges to QM with a clean first-order law, $\max_\tau|\Delta P| = 1.24/n$ for $|1,1\rangle$ and $1.02/n$ for $|2,0\rangle$ (verified numerically to $n = 512$), rising to $3.8/n$ for four-photon inputs; the $8$--$13\%$ residual that remains under the single-rotation calibration (Fig.~\ref{fig:scaling}) is reabsorbed into the angle axis by the refined dictionary.
The measured states are degenerate in $\ell_{a1}$, which is why it must be summed: the refined model makes exactly \emph{one} prediction per $n$, and $n$ acquires the physical reading of an information capacity.
The composition-consistent model is therefore the primary target of this work: it is the unique formulation with a quantum limit, since under the single-rotation calibration the deviation does not vanish but grows with $n$ toward a plateau of $\approx0.13$.
The published linear variants remain the concrete hypotheses in print, and the same data exclude them outright; we report those constraints for completeness.

Written in counts, the composition law $z_{12} = z_1 z_2$ reads $\widetilde{A}_{12} =
\widetilde{A}_1\widetilde{A}_2 - \widetilde{B}_1\widetilde{B}_2$,
$\widetilde{B}_{12} = \widetilde{A}_1\widetilde{B}_2 +
\widetilde{B}_1\widetilde{A}_2$: coincident flips \emph{interfere
destructively}, with the same sign alternation the model assigns to its path quantum numbers \cite{StojkovicNote2026}.
The composed map lives at $n_{12} = n_1 n_2 - 2\widetilde{B}_1\widetilde{B}_2$: cascaded rotations occupy a much finer grid, suppressing discreteness to $O(1/n_1 n_2)$.
We verified both the algebra and the probability-level consistency of the composed maps numerically.
The alternative rule, naive position-wise composition in which coincidences \emph{add}, under-rotates, and fed through the calibration it predicts strongly split-dependent outcomes when one total rotation is implemented as two partial ones.
Cascaded beam splitters therefore discriminate the two composition laws directly, independently of the finite-$n$ deviation itself.

\begin{figure}[t]
\includegraphics[width=\columnwidth]{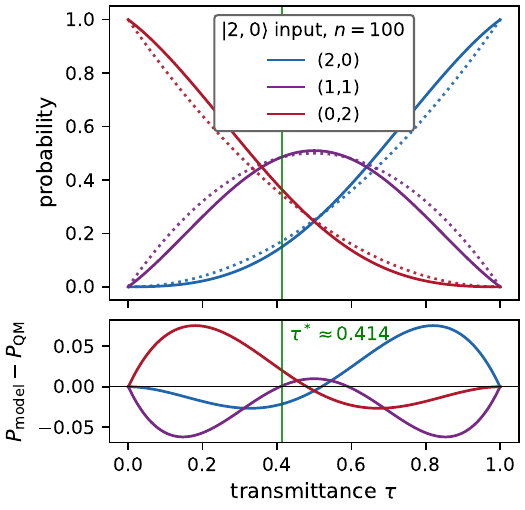}
\caption{Sequence-model (solid) and QM (dotted) partition probabilities
for a $|2,0\rangle$ input at $n=100$, with their difference below.
The green line marks the null-differential working point $\tau^* = 0.414$ where the model and QM agree for the $(1,1)$ outcome while disagreeing for $(2,0)$ and $(0,2)$.}
\label{fig:model}
\end{figure}

\begin{figure}[t]
\includegraphics[width=\columnwidth]{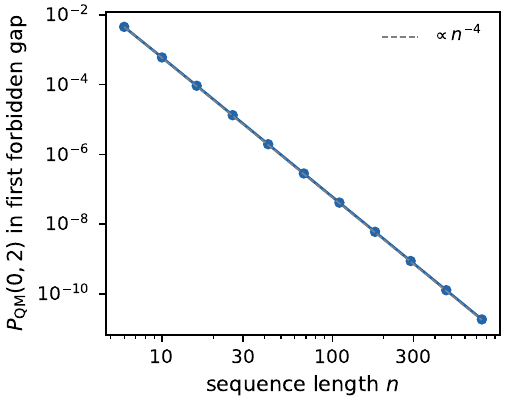}
\caption{Granularity signature: the QM probability of the outcome that
the finite-$n$ model forbids in the last rational grid gap $\tau \in (\cos^2(\pi/2n),\,1)$ falls as $n^{-4}$.}
\label{fig:forbidden}
\end{figure}

In the published parametrization the treatment of $\ell_{a1}$ defines a family of variants (Fig.~\ref{fig:scaling}).
If $\ell_{a1}$ is summed as an internal nuisance (the unconditioned variant), the deviation from QM \emph{plateaus} at the $8\%$ level, essentially independent of $n$.
If a single $\ell_{a1}$ may be chosen per experiment, the deviation falls like $1/n$ down to a $\approx 1\%$ floor.
Only if $\ell_{a1}$ may be retuned for every measurement setting does the deviation follow a clean $1/n$ law, $\max_\tau|\Delta P| \approx 0.63/n$.
The composition-consistent parametrization resolves this ambiguity from within the theory: $\ell_{a1}$ is summed, and the plateau is revealed as an artifact of the linear angle dictionary.
The published variants are the predictions available to the community in print; we therefore test them independently of the refinement, and then test the refined model as well.
Two further signatures matter for experiment design: the deviation is maximal for four-photon inputs (non-monotonic in photon number), and near $\tau \to 0,1$ the rational grid forbids outcomes to which QM assigns probability $\propto n^{-4}$ (Fig.~\ref{fig:forbidden}), so a single observed ``forbidden'' event is evidential without any distribution fitting.

\begin{figure}[t]
\includegraphics[width=\columnwidth]{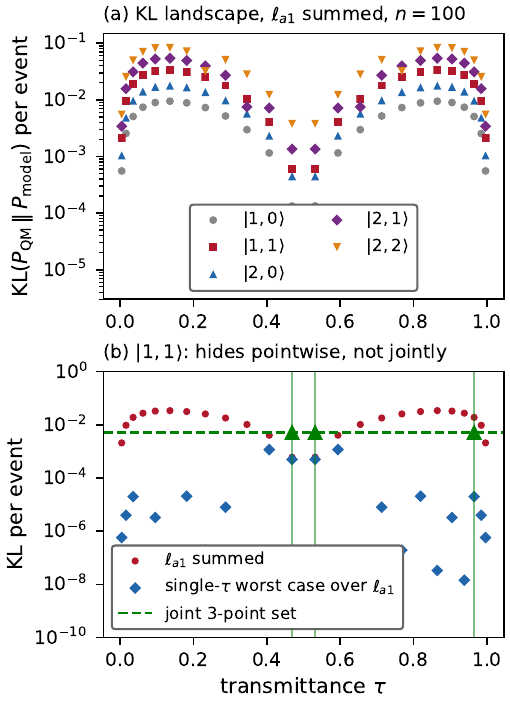}
\caption{Experiment-design optimization. (a) Discrimination landscape:
per-event KL divergence between QM and the unconditioned model ($n=100$) for candidate input states.
(b) For $|1,1\rangle$, the conditioned model evades any single-$\tau$ test by tuning $\ell_{a1}$ (dots), but no single $\ell_{a1}$ fits a set of transmittances jointly: the optimal 3-point set (triangles, dashed level) restores percent-scale discriminability.}
\label{fig:optimizer}
\end{figure}

Because the conditioned variants can adapt $\ell_{a1}$, measurement settings must be chosen so that no single internal choice can reproduce QM everywhere.
We select settings by maximizing the per-event Kullback--Leibler divergence ${\rm KL}(P_{\rm QM}\,\|\,P_{\rm model})$, which is the exponent of the likelihood-ratio test, over input states and grid transmittances, in the worst case over $\ell_{a1}$ (Fig.~\ref{fig:optimizer}).
The native two-photon input $|1,1\rangle$ is the most cost-effective discriminator (no heralding overhead, and the interference sector of the model deviates most): roughly 330 accepted events per point suffice for a $5\sigma$ separation from the unconditioned variant.
And while the conditioned model can hide at any \emph{single} transmittance (worst-case KL collapses below $10^{-6}$), one $\ell_{a1}$ cannot fit a \emph{set} of transmittances jointly: the optimal three-point set $\tau \in \{0.469, 0.531, 0.965\}$ restores a worst-case KL of $5.1\times10^{-3}$ per event, i.e.\ $\sim$2\,500 events for $5\sigma$.
This experiment-design step, physics-informed optimal design in the sense of adaptive Bayesian experiment design
\cite{Fiderer2021,Bavaresco2024,Cortes2022,Valeri2020}, is what makes
the conditioned variants testable at all.
For the composition-consistent model the internal freedom disappears and the design problem reduces to maximizing information about $n$ against the platform systematics; the multi-setting full-curve design remains optimal in that limit.

\section{Platform and protocol}
\label{sec:protocol}

Measurements were performed on Quandela's cloud QPUs: Ascella
\cite{Maring2024ascella} (12-mode platform built on a demultiplexed
quantum-dot single-photon source \cite{Arakawa2020,Flagg2012,Eisaman2011} with superconducting-nanowire detection \cite{EsmaeilZadeh2021}; June 2025 dataset) and its successor Belenos (24 modes, up to 12 photons; 2026 campaigns), programmed through the Perceval framework
\cite{Heurtel2023}. These processors belong to the linear-optical
architecture class established by on-chip boson sampling
\cite{Spring2013,Wang2019boson}. We emphasize the mode of use: the
processors are not employed to simulate the model or to emulate some other system.
The variable beam splitter under test \emph{is} the physical interferometer chip itself, and every data point in this paper is a direct measurement of single photons propagating through it.
Simulation enters this work only in the offline validation of the analysis pipeline (Sec.~\ref{sec:pipeline}).
The platform imposes two constraints that shaped the protocol.
First, multi-photon input modes are not supported, so a $|2\rangle$ Fock input cannot be programmed directly: we prepare it by interfering the native $|1,1\rangle$ input on a balanced coupler (Hong--Ou--Mandel bunching \cite{HOM1987,Sturges2021}) and post-selecting on an empty herald mode, an end-of-circuit selection that runs unmodified on hardware with acceptance $\tfrac12$.
Second, the detector model must be verified rather than assumed: the platform metadata declare threshold detectors, for which two photons in one mode would register a single click and a splitter-tree readout with an exact factor-2 acceptance correction would be required.
Empirically, returned Belenos states contain multi-photon mode occupations, i.e.\ the readout is photon-number resolving; our pipeline auto-detects the readout class from the data and applies the correction only when warranted.

The campaigns interleave five job classes: (i) $|1,1\rangle$ partition measurements at the optimized transmittances and, in the second campaign, across a 25-point grid covering the full Hong--Ou--Mandel curve; (ii) $|1,0\rangle$ single-photon anchors at the same settings, measuring the intensity calibration $\hat\tau(\tau_{\rm nom})$; (iii) repeated reference points at $\tau = 0.5$, tracking source stability during the campaign; (iv) heralded $|2,0\rangle$ partitions at the null-differential point $\tau^*$ and companions; (v) in the third (composition) campaign, cascaded pairs ${\rm BS}(s\,\theta_{12})\,{\rm BS}((1{-}s)\,\theta_{12})$ at splits $s \in \{0.1, 0.25, 0.5\}$ alongside single-BS references, with compiled circuits resubmitted in up to four time-separated rounds; this class serves simultaneously as a composition-law test and as a direct measurement of the compilation reproducibility.
A zero-reflectance job measures the dark- and background-count floor (terminology follows Ref.~\cite{Bienfang2023}).
All raw counts, job identifiers, and per-job source performance metrics are logged for provenance; every analysis in this paper runs from those records.

\section{Analysis pipeline and validation}
\label{sec:pipeline}

Counts are reduced to partition fractions with heralding filters and compared to hypotheses through joint multinomial likelihoods.
The null hypothesis is QM dressed with the in-situ nuisance chain: a quadratic map $\hat\theta(\theta_{\rm nom})$ for the programmed beam-splitter angle plus an interference-specific offset; the source indistinguishability $V$ \cite{Grice1997}, entering as the standard mixture $P = V P_{\rm indist} + (1{-}V) P_{\rm dist}$; a time-linear source stability term $V(t)$ constrained by the interleaved references; and the output-port efficiency ratio $\varepsilon = \eta_1/\eta_2$ reweighting partitions as $\varepsilon^{k_1}$.
The alternative hypotheses replace the indistinguishable-sector prediction with the sequence model at given $n$ (and, for the conditioned variant, profiled $\ell_{a1}$), inheriting the same nuisances.
Confidence statements use the likelihood ratio $-2\Delta\ln L$; for the published-variant exclusions, which exceed the asymptotic $3.84$ (95\%, 1 dof) threshold by one to three orders of magnitude, the precise calibration is not limiting.
For the composition-consistent comparison neither hypothesis nests the other, so no asymptotic threshold applies and the calibration is by parametric bootstrap under both hypotheses (Sec.~\ref{sec:bounds}).

For the final constraints on the composition-consistent model the dominant platform systematic is promoted from a scatter estimate to a random effect: each programmed circuit carries a latent compiled-angle offset $\delta_i \sim \mathcal{N}(0, \sigma_\theta^2)$, marginalized point-by-point on a dense grid, with $\sigma_\theta$ profiled as a nuisance.
The per-point likelihood in $\delta_i$ is an order of magnitude narrower than the offset prior, so fixed-node quadrature undersamples it; the dense marginal was verified by step-halving and against a per-point Laplace control.
The fit returns $\sigma_\theta = 0.078$ ($\sigma_\tau \approx 0.039$ at $\tau = 0.5$), the same scale as the residual overdispersion of the full curve ($\sigma_\tau = 0.026$) and compatible with the frozen-offset picture established by the circuit-repetition measurement of Sec.~\ref{sec:results}; the excess over the pure angle-jitter estimate absorbs residual structure beyond compilation jitter, and the sensitivity checks of Sec.~\ref{sec:bounds} show the resulting bound does not depend on it.
This marginalization removes the sign instabilities that plague the raw likelihood ratio once the model deviation falls below the jitter scale.

The pipeline was validated end-to-end before any hardware use: on ideal local simulation all three circuit classes reproduce the quantum predictions with uniform $\chi^2$ $p$-values across full transmittance sweeps; the heralded-$|2\rangle$ construction reproduces the Fock binomial exactly with acceptance $0.500$; and a synthetic threshold-detector collapse test closes the pseudo-photon-number correction.
A dedicated noise study established which imperfections can and cannot mimic the model: source distinguishability, $g^{(2)}$ contamination, and \emph{symmetric} loss leave the same-port partition observable exactly on the quantum binomial (loss commutes with the beam splitter; photon-number post-selection removes it), whereas
\emph{asymmetric} detector efficiency shifts the partition ratio by
amounts comparable to the model signal but with a smooth, antisymmetric $\tau$-shape that is cleanly distinguishable from the model's oscillatory signature and is calibrated out by the single-photon anchors.
The same conclusion holds in an exact Fock-space simulation of a squeezed-light/photon-number-resolving platform (Xanadu-class), establishing that the protocol transfers across photonic architectures.

\section{Hardware results}
\label{sec:results}

\begin{figure}[t]
\includegraphics[width=\columnwidth]{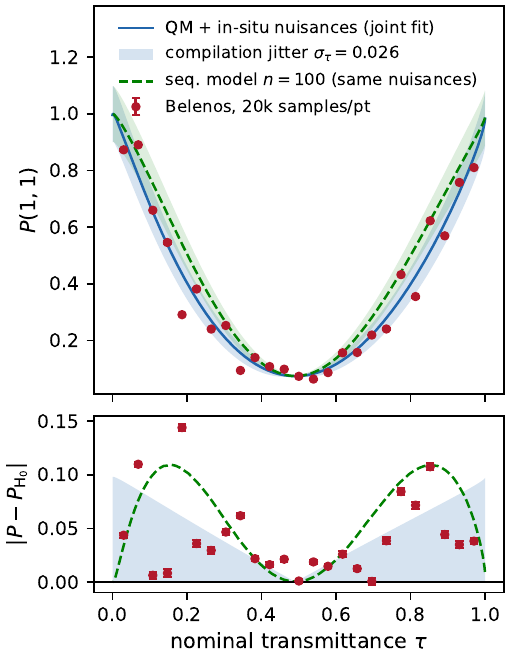}
\caption{Belenos full-curve campaign: $P(1,1)$ for the native
$|1,1\rangle$ input across 25 transmittances (20\,000 samples each; statistical error bars are smaller than the symbols), against QM with jointly fitted in-situ nuisances (blue) and the unconditioned sequence model at $n=100$ under the same nuisances, in the published linear parametrization, evaluated on its native rational transmittance grid (green).
Shaded bands show the per-job compilation reproducibility ($\sigma_\tau = 0.026$, measured from the scatter of the run itself) around each hypothesis; the lower panel shows the magnitude of the residuals with the same band around QM.
The model requires a coherent $+0.10$--$0.12$ excess at both flanks; the data scatter symmetrically about the quantum curve.}
\label{fig:v2}
\end{figure}

\begin{figure}[t]
\includegraphics[width=\columnwidth]{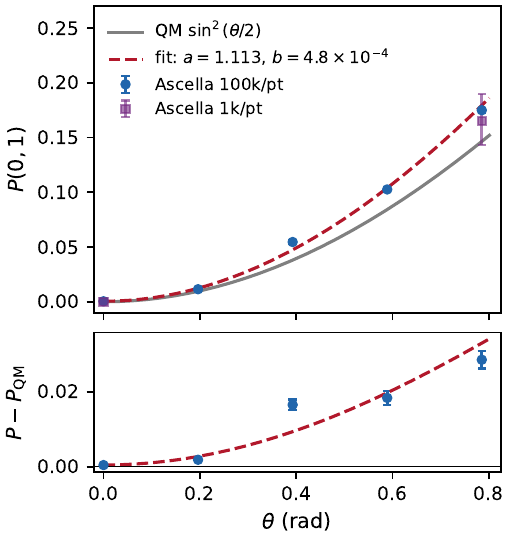}
\caption{Ascella $|1,0\rangle$ sweeps with
95\% credible intervals, against ideal QM and a two-parameter calibration fit (angle scale $+$ incoherent floor).}
\label{fig:reanalysis}
\end{figure}

\emph{Baseline (Ascella, 2025).} The $|1,0\rangle$ sweeps
($5\times10^5$ events) show the measured reflectance exceeding the ideal $\sin^2(\theta/2)$ with overwhelming statistical significance (binomial $p$-values reaching $10^{-146}$).
The excess is entirely attributable to an angle recalibration $a = 1.11$ and an incoherent floor $4.8\times10^{-4}$ anchored by the $\theta=0$ blind point (Fig.~\ref{fig:reanalysis}).
This motivates the central design rule of the campaign: absolute transmittances are calibration-limited at the percent level, so model tests must either fit the calibration jointly or use metrics from which it cancels.
Even so, granting the model its own best-fit calibration, the unconditioned variant fits these data worse than calibrated QM by $2\Delta\ln L = 791$--$1758$ for every $n$ from 16 to 512 (Fig.~\ref{fig:exclusion}): because its predicted deviation does not shrink with $n$, this variant is excluded outright, for all $n$, by the single-photon data alone.

\begin{figure}[t]
\includegraphics[width=\columnwidth]{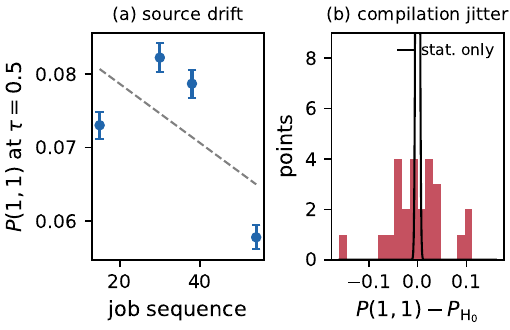}
\caption{In-situ systematics of the Belenos campaign. (a) Repeated
$\tau=0.5$ references reveal source stability variation over the campaign, absorbed by a time-dependent indistinguishability nuisance.
(b) Distribution of point residuals against the fitted null hypothesis: compilation jitter broadens the scatter twenty-fold beyond the statistical width (black, $\sigma = 0.003$).}
\label{fig:systematics}
\end{figure}

\emph{Full-curve campaign (Belenos, 2026).} The main campaign comprises
58 jobs and $\sim 9\times10^5$ accepted events; the data follow the quantum prediction across the entire Hong--Ou--Mandel curve (Fig.~\ref{fig:v2}).
The nuisance chain returns $V = 0.85$, $\varepsilon = 1.10$, a leakage floor of $2.5\times10^{-4}$, and a smooth quadratic angle calibration; with the full curve constraining the interference phase, the phase-specific offset collapses to $10^{-3}$.
The interleaved references reveal source stability variation of $\Delta P(1,1) \approx 0.02$ over the $\sim$30-minute campaign [Fig.~\ref{fig:systematics}(a)], absorbed by the $V(t)$ nuisance.
The dominant residual is a compilation-to- compilation transmittance jitter: point residuals scatter with $\sigma_P = 0.055$, twenty times the statistical error, with no positive correlation between settings [Fig.~\ref{fig:systematics}(b)].
Decomposing the residual variance into its statistical and jitter parts, $\langle r^2\rangle = \langle\sigma_{\rm stat}^2\rangle +
\langle(\partial P/\partial\tau)^2\rangle\,\sigma_\tau^2$, yields
$\sigma_\tau = 0.026$ per programmed circuit, measured from the scatter of the run itself; the shaded bands in Fig.~\ref{fig:v2} show this reproducibility envelope around both hypotheses.

\begin{figure}[t]
\includegraphics[width=\columnwidth]{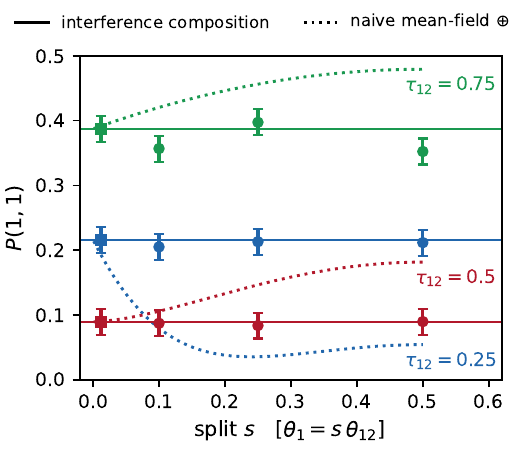}
\caption{Composition-law test. Cascaded implementations
${\rm BS}(s\,\theta_{12})\,{\rm BS}((1{-}s)\,\theta_{12})$ of three total transmittances against the split $s$: measured $P(1,1)$ (circles; squares are the single-BS anchors) is split-invariant, as demanded by the interference composition $z_1 z_2$ (solid lines at the anchor values).
The naive position-wise composition (dotted), anchored differentially at $s \to 0$, predicts shifts the data exclude at $\chi^2 = 98.5$ for $9$ degrees of freedom ($8\sigma$).
Error bars: frozen per-circuit systematic ($0.02$) and statistics in quadrature.}
\label{fig:composition}
\end{figure}

\emph{Composition campaign (Belenos, 2026).} The third campaign
implements the same total rotations $\tau_{12} \in \{0.25, 0.5, 0.75\}$ as cascades of two partial rotations at three splits, plus single-BS references, with compiled circuits resubmitted in four rounds (41 completed jobs, two to four repeats per circuit); the first three rounds ran on a single day, and five circuits were remeasured eight weeks later, after an intervening platform maintenance.
First, the nature of the compilation jitter: within the single-day set, repeated submissions of the \emph{same} compiled circuit reproduce to $0.003$--$0.019$ in $P(1,1)$ (median $0.010$), well below the circuit-to-circuit scatter $\sigma_P = 0.055$ of the full-curve campaign, and different circuits miss the prediction calibrated on the full-curve campaign by up to $0.1$: the compilation offset is frozen per circuit rather than random per run, and is therefore calibratable.
The delayed remeasurement qualifies the freezing timescale: all five circuits shift coherently downward, by $-0.04$ to $-0.08$ for the four $\tau_{12}=0.75$ circuits and by $-0.02$ for the one $\tau_{12}=0.25$ cascade, so the offsets are stable within a platform calibration epoch but not across recalibrations, and a calibrate-then-measure protocol must complete within one epoch.
Second, the composition law: the measured outcomes are split-invariant (median deviation from the single-BS anchors $0.006$, at most $0.04$, within the differential errors; Fig.~\ref{fig:composition}), where the naive count-composition rule predicts split-dependent shifts of up to $0.2$ in $P(1,1)$.
A calibration-free differential test against the single-BS anchors, with the frozen per-circuit floor of $0.02$ charged independently to every circuit and the shared-anchor correlations propagated, excludes the naive rule at $\chi^2 = 98.5$ for $9$ degrees of freedom, an $8\sigma$ Gaussian equivalent, while the interference-sign composition fits the same data at $\chi^2 = 3.9$.
The exclusion is not floor-limited: it remains above $5\sigma$ if the floor is raised to $0.026$, the cross-circuit value of the full-curve campaign.
The residual scatter at the Hong--Ou--Mandel dip, where the transmittance sensitivity vanishes, is source-stability variation ($\Delta V \approx 0.03$), consistent with the interleaved references.

\section{Constraints and platform sensitivity benchmark}
\label{sec:bounds}

\begin{figure}[t]
\includegraphics[width=\columnwidth]{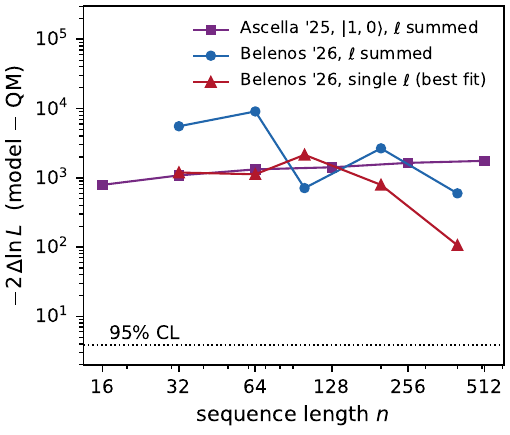}
\caption{Model exclusion across platforms and variants: likelihood ratio
of the sequence model to QM, both dressed with the full in-situ nuisance chain (drift-corrected for Belenos), against sequence length.
All points lie one to three orders of magnitude above the 95\% threshold (dotted).}
\label{fig:exclusion}
\end{figure}

For the composition-consistent parametrization, the primary target, the random-effects likelihood of Sec.~\ref{sec:pipeline} is the appropriate instrument, and it yields the central bound of this work.
Because neither hypothesis nests the other, each $-2\Delta\ln L(n)$ is calibrated by parametric bootstrap under both hypotheses: the full campaign is regenerated from the fitted null and from the fitted model at that $n$, with all nuisances refitted per replicate ($B = 150$).
The observed sequence decays from $15023$ at $n = 8$ through $3463$ at $16$, $1211$ at $32$, $413$ at $48$, $202$ at $64$, and $81$ at $100$ to $-17$ at $200$ and $-12$ at $400$.
For $n \le 100$ the observed values lie far outside the model-truth bands: at $n = 100$ the band under the fitted model has mean $-49$ and 99th percentile $-20$, twelve band widths below the observed $+81$; all $n \le 100$ are excluded far beyond the 99\% bootstrap level.
At $n \ge 200$ the discrimination power of the campaign is exhausted: the observed value falls within the model-truth band at $n = 200$ and slightly below both bands at $n = 400$, a negative excess reflecting residual time structure rather than model preference.
Two sensitivity checks support this reading: modeling the measured source drift as $V(t)$ in both hypotheses moves the statistic from $-17$ to $-11$ at $n = 200$ and from $-12$ to $-6$ at $n = 400$ while leaving the $n = 100$ exclusion intact ($+51$), and replacing the Gaussian offset distribution by a Student-$t(4)$ changes all values by at most $0.2$.
We therefore quote $n > 100$ and make no statement beyond $n \approx 150$.
Stated physically: \emph{the information capacity of the register carrying the two-photon state, if finite, exceeds} $\sim\!10^2$.
The core of this number is close to model independent.
Any partition law that is constant on at most $n+1$ cells of the setting axis must miss the quantum curve, whose total variation in the $P(1,1)$ channel is $2$ over the sweep, by at least $1/(n+2)$ somewhere; the sequence model's $1.24/n$ sits only $25\%$ above that floor.
The bound therefore probes finite resolution itself, with the model supplying the grid and levels the likelihood analysis tests.
The floor does not by itself transfer the exclusion to the whole class: a staircase with free cell boundaries and levels can interpolate any finite set of programmed settings, so its deviation can hide between the measured points.
What no cell assignment evades is a slope measured \emph{inside} a cell, which is why the differential test of Sec.~\ref{sec:outlook} is the class-wide version of this bound.
The cascade campaign contributes the complementary, deviation-independent statement that the composition sector of the model behaves quantum mechanically (Fig.~\ref{fig:composition}).

The published linear-parametrization variants are the easily excluded alternative, and Fig.~\ref{fig:exclusion} collects their constraints; every test grants the model full benefit of the doubt, with its own best-fit nuisance chain and free internal parameters.
The unconditioned variant is excluded on both processor generations and both input classes, an architecture-independent result; because its predicted signal is $n$-independent, the exclusion holds for all sequence lengths.
The conditioned (single-$\ell_{a1}$) variant, testable only through the jointly designed multi-setting campaign, is excluded for $n = 32$ to $400$ with drift-corrected margins between $-2\Delta\ln L = 106$ and $2148$, far beyond any plausible calibration uncertainty of the raw ratio.
Below $n = 32$ this variant is untested rather than excluded: the multi-setting grid began at $32$, and the variant's predicted deviation only grows toward small $n$.
The per-setting-conditioned variant remains essentially unconstrained: its $0.63/n$ signal falls below the current sensitivity.

\begin{figure}[t]
\includegraphics[width=\columnwidth]{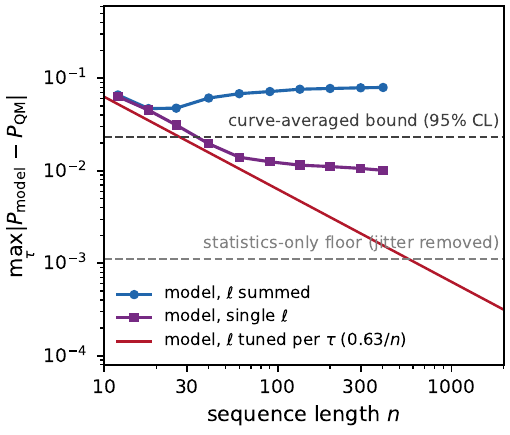}
\caption{Sensitivity ladder. Model deviation scales for the three
variants against sequence length, compared with the measured 95\% curve-averaged coherent-deviation bound of this work (dashed; a bound on the curve average, not on the plotted maximum) and the statistics-only floor of the same campaign if compilation jitter were eliminated.
The single-$\ell$ variant is nonetheless excluded to $n=400$ by the shape-based likelihood test, which uses all outcomes and datasets jointly rather than a single amplitude; the composition-consistent model ($1.24/n$ for $|1,1\rangle$) lies a factor two above the lowest line and is bounded to $n > 100$ by the bootstrap-calibrated random-effects analysis.}
\label{fig:sensitivity}
\end{figure}

Model-independently, the full-curve campaign bounds a \emph{coherent} (curve-averaged) deviation from the quantum partition law in the $P(1,1)$ channel at $|\overline{\Delta P}| < 2.3\times10^{-2}$ (95\% CL; unweighted mean of the 25 post-fit residuals with a Student-$t$ interval): compilation jitter averages down over the curve while a common-mode deviation would not.
The residuals show no positive inter-setting correlation (lag-1 autocorrelation $-0.6$).
Two qualifications delimit what this bounds.
It constrains the component of a deviation orthogonal to the in-situ calibration chain; a deviation degenerate with the fitted nuisances, most notably the indistinguishability $V$, is absorbed by construction, and only an externally anchored $V$ or a cross-architecture comparison removes that degeneracy.
And it is a bound on the curve average, not on $\max_\tau|\Delta P|$; the sequence model's own oscillatory deviation shape averages close to zero over the grid, so the capacity bounds on $n$ come from the likelihood analysis above, not from this number.
Figure~\ref{fig:sensitivity} places this bound, and its foreseeable successors, against the model's deviation scales.
The statistical floor of the same dataset is $1.1\times10^{-3}$; compilation jitter costs a factor twenty and is not fundamental: the composition campaign demonstrates that the offset is frozen per compiled circuit within a platform calibration epoch, so a calibrate-then-measure protocol that completes on one epoch with a fixed circuit removes it, opening the $10^{-3}$ floor ($n \sim 10^3$) to current cloud hardware.
Four-photon inputs would triple the per-event signal ($3.8/n$), but at the measured Belenos rates a single accepted $|2,2\rangle$ event costs several hundred thousand source shots, two orders of magnitude more machine time per unit sensitivity than the two-photon protocol, so the two-photon classes remain the cost-effective discriminators on the current platforms.
The present limits are those of the cloud platform, not of the observable: beam-splitter photon statistics are a mature metrological tool \cite{Lyons2018,Ndagano2022}.

\section{Outlook}
\label{sec:outlook}

Three extensions follow directly.
A calibrate-then-measure campaign, which maps the frozen offset of a handful of compiled circuits and then accumulates statistics on those same circuits within one calibration epoch, would promote the present bounds to the $10^{-3}$ statistical floor and extend the information-capacity limit toward $n \sim 10^3$.
That range contains a concrete physical target: under the Bekenstein reading of $n$, the fiber-like modes of the Belenos interferometer itself correspond to $n_{\rm Bek} \approx 3\times10^{2}$, a factor three beyond the present bound, so the capacity hypothesis becomes testable at the platform's native geometry.
Beyond that range a dedicated laboratory campaign at larger $n$ is the natural continuation, pairing photon-number-resolved source characterization \cite{Thomay2017} and statistically background-free single-photon detection \cite{Cheng2015}, which the forbidden-outcome search requires, with machine-learning classification of higher-order photon states, which reaches the required event rates in simulation \cite{Xu2024}; its quantitative design is beyond the scope of this paper.
The same protocol transfers to squeezed-light platforms with native photon-number-resolving detection \cite{Madsen2022borealis,Heinzel2026}, where post-selection on total photon number prepares the $|2\rangle$ sector exactly and symmetric loss provably leaves the partition observable unbiased; a cross-architecture deviation, were one ever observed, is the signature that would distinguish physics from platform.
On the theory side, the composition-consistent parametrization elevates $n$ from a model parameter to a measurable, the information capacity of the register behind a prepared state, and extends naturally beyond rotations: boosts, as non-compact partners of rotations, compose natively in the counts without an interference sign \cite{StojkovicNote2026}, suggesting that the same experimental program has a special-relativistic sector.
The most valuable remaining theoretical input is a closed-form large-$n$ asymptotic of the model, without which predictions at $n \gtrsim 10^4$ are computationally out of reach.

\begin{acknowledgments}
We thank S.~Powers for the reference implementation of the model.
This work was supported in part by the National Science Foundation (EAGER award No.~2533850).
Cloud hardware access was provided in part through Quandela's academic program.
\end{acknowledgments}

\section*{Data availability}
The raw measurement records of all campaigns and the complete analysis code will be made available on Zenodo at \href{https://doi.org/10.5281/zenodo.22676818}{doi:10.5281/zenodo.22676818} upon publication.

\bibliography{refs}

\end{document}